\documentclass[structabstract]{aa}
\usepackage{txfonts}
\usepackage{graphicx}
\usepackage{longtable}
\usepackage{supertabular}
\usepackage{lscape}
\usepackage{hhline}
\usepackage{setspace}
\usepackage{threeparttable}
\usepackage{pifont}

\begin{document}
\title{A first study of the galaxy HRG\,2304 and its companion AM\,1646-795 (NED01) \thanks{
Based on  observations performed with the telescope of 3.6m at the European Southern Observatory (ESO), La Silla, Chile.}}

\titlerunning{The galaxy HRG\,2304 and its companion}
\authorrunning{Wenderoth et al.}

\offprints{Max Fa\'undez-Abans; max@lna.br}

\author{E. Wenderoth
          \inst{1}
          \and
          M. Fa\'undez-Abans\inst{2}\and A.C. Krabbe\inst{3}\and M. de Oliveira-Abans\inst{2}\and H. Cuevas\inst{4} 
          }

   \institute{Gemini Observatory,
              Southern Operation Center, c/o AURA, Casilla 603, La Serena, Chile.\\
              \email{ewenderoth@gemini.edu}
         \and
             Laborat\'orio Nacional de Astrof\'isica, Rua Estados Unidos 154, 37504-364, Itajub\'a, MG, Brazil.\\
             \email{max@lna.br; mabans@lna.br}
         \and
             Universidade do Vale do Para\'{i}ba - UNIVAP. Av. Shishima Hifumi, 2911 - Urbanova
    CEP: 12244-000 - S\~{a}o Jos\'{e} dos Campos, SP, Brazil.
             \email{angela.krabbe@gmail.com}
          \and
             Universidad de La Serena, Benavente 980, La Serena, Chile.\\
             \email{hcuevas@dfuls.cl}
             }

\date{Received 24 February 2011 / Accepted 21 March 2011}


\abstract
   {}
{We report the first study of the peculiar ring-like galaxy HRG\,2304 (NED02),
which   was previously classified as a ring galaxy with an elliptical smooth ring. This object was selected to prove that it is a candidate for the Solitaire-type ring galaxies in an early stage of ring formation. The main goal of this work is to provide the spectral characteristics of the current object and its companion AM\,1646-795 (NED01).}
   {The study is based on spectroscopic observations in the optical band to highlight the characteristics of this interacting galaxy. To investigate the star formation history of HRG\,2304 we used the stellar population synthesis code STARLIGHT. 
   The direct V and B broad band images were used to enhance some fine structures.}
   {Along the entire long-slit signal, the spectra of HRG\,2304 and its companion resemble that of an early-type galaxy. We estimated a heliocentric systemic redshift of z = 0.0415, corresponding to heliocentric velocities of 12\,449 km s$^{-1}$ for HRG\,2304 (NED02) and 12\,430 km s$^{-1}$ for AM1646-795 (NED01). The spatial variation in the contribution of the stellar population components for both objects are dominated by an old stellar population $2\times10^{9} <\rm t \leq 13\times10^{9}$ yr. The observed radial-velocity distribution and the fine structures around HRG\,2304 suggest an ongoing tidal interaction of both galaxies.}
   {The spectroscopic results and the morphological peculiarities of HRG\,2304 can be adequately 
   interpreted as an ongoing interaction with the companion galaxy. Both galaxies are early-type, the companion is elliptical, and the smooth distribution of the material around HRG\,2304 and its off-center nucleus in the direction of AM1646-795 (NED01) characterize HRG\,2304 as a Solitaire-type galaxy candidate in an early stage of ring formation.}


   \maketitle
%

\section{Introduction}

In the past decades, studies have shown that the gravitational interaction is the most 
important factor in galactic evolution. Gravitational interaction  directly affects properties 
such as size, morphological type, luminosity, star formation rate (SFR), and mass distribution 
in galaxies. Gravity is the acting force that makes galaxies interact, collide, and merge. 
Colliding galaxies are twisted and deformed by their mutual gravitational fields. This often 
gives galaxies peculiar shapes and can create tails, plumes, and bridges between the interacting objects. 
These features are the patterns of the gas, dust, and some stars, which have been drawn out by 
tidal forces during  interaction. In addition, interactions are seen in galaxies that form rings. 
An intruder galaxy can plunge through the center (or close to it) of a rotating disk galaxy, triggering the birth of 
bright young stars in the wake of radially expanding annular shock waves. Those objects are called ring galaxies (RGs) 
and belong to the morphological category 6 of the ``Catalogue of Southern Peculiar Galaxies and Associations"  
(see Arp \& Madore \cite{am1977}, \cite{am1986}, as a starting point in the study of peculiar galaxies with rings).
In general, the rings may either be collisional (peculiar -- pRG) or resonant 
(normal -- NRG, see Fa\'{u}ndez-Abans \& de Oliveira-Abans \cite{foa98a} - hereafter FAOA, for this denomination). 
It is well accepted that collisional RGs are important regions for studying disks that remain fairly intact after the 
interaction, allowing detailed observation of their structures and star formation properties, such as galaxy-scale 
perturbation properties, hydrodynamics, vigorous non-nuclear star formation, and stellar evolutionary processes across
and inside the ring. A description of morphological characteristics of \emph{pRG} is found in FAOA. General properties 
of those galaxies can be found in Theys \& Spiegel (\cite{ts76,ts77}), Lynds \& Toomre (\cite{lt76}), 
Appleton \& Struck-Marcell (\cite{app96}), and for a short review see Dennefeld \& Materne (\cite{denn80}).

In the work on the morphology of pRGs by FAOA, the morphological categories are compressed into five families, 
following the general behavior of the galaxy-ring structures. Eight morphological subdivisions are highlighted 
in Table 1 of that paper. One of these morphological subdivisions is a basic structure called ``Solitaire".
The pRG Solitaire is described as an object with the bulge on the ring, or very close to it, resembling a 
one-diamond finger ring (single knotted ring). In these objects, the ring generally looks smooth and almost thin on
the opposite side of the bulge (as archetypes see FM\,188-15/NED02 and AM\,0436-472/NED01). In spite of the poor statistics, the companion galaxy may be almost elliptical. In Fa\'{u}ndez-Abans et al. 
(\cite{fa2011}), a preliminary study and statistics of solitaire-like objects are carried out with the existing 
cataloged pRG-solitaire type objects. 

No interacting galaxies in the early stage of formation of a Solitaire-type morphology are reported in the literature, so the galaxy \object{HRG\,2304} was selected as a candidate. In this paper, we report the first results of a  
study of the tidally disturbed galaxy HRG\,2304 (NED02) and its companion AM\,1646-795 (NED01) based on data 
obtained from long-slit spectrophotometric observations, at ESO-La Silla Observatory, in Chile.

\section{Early data and the field around HRG\,2304}

The galaxy HRG\,2304 looks like an elliptical galaxy on the Digitized Sky Survey (DSS) (see also: the film copies 
of the ESO/UPPSALA survey of the ESO(B) atlas, FAOA and NED for more references). As previously mentioned, no dedicated study has been done on this object and it appears only 
in a few references: (a) in a list of probable ring galaxies in a limiting magnitude of 
17.5 in the J band (Fa\'undez-Abans, Cuevas and Hertling, \cite{fch}); (b)
in a list of morphological characteristics of pRG by FAOA; (c) in the 2MASS-selected 
flat galaxies catalog (Mitronova et al. \cite{mitro}); and (d) in the catalog of near-infrared 
properties of LEDA galaxies by Paturel et al. (\cite{pat}).

\begin{figure}
   \centering
\includegraphics[width=8 cm]{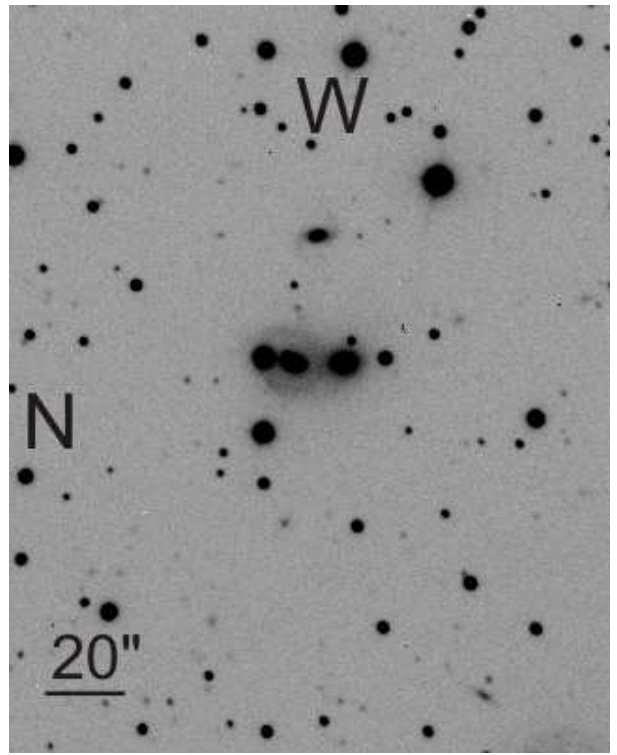}
   \caption{Field around HRG\,2304 in the V band. West is up and north is to the left. 
   There are two prominent objects aligned from N to S, encompassing HRG\,2304, from left to  
   right: the star GSC094100228 of 14.26 V mag, HRG\,2304, and the AM\,1646-795 (NED01) galaxy.}
    \label{Fig.1}
    \end{figure}

\begin{figure}
   \centering
\includegraphics[width=8 cm]{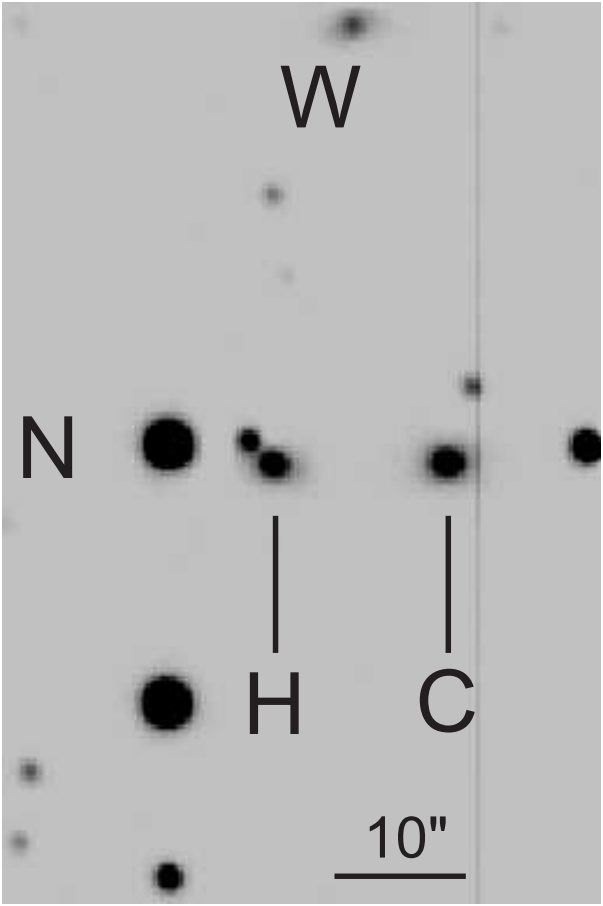}
   \caption{A zoom of the field around HRG\,2304 in the V band.
   West is up and north is to the left. 
   There is a star projected on the HRG\,2304 galaxy near the bulge, N-W direction. 
   The image was enhanced using a median filtering of 30 square-pixel kernel (Fa\'undez-Abans \& de Oliveira-Abans\cite{foa98b}). 
   The H letter indicates HRG\,2304 (NED02) and C is AM\,1646-795 (NED01), the companion.}
    \label{Fig.2}
    \end{figure}

\begin{figure}[ht]
\begin{center}$
\begin{array}{cc}
\resizebox{\hsize}{!}{\includegraphics[]{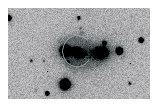}}\\
\resizebox{\hsize}{!}{\includegraphics[angle=0]{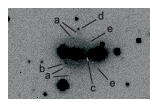}}
\end{array}$
\end{center}
\caption{Some structures are highlighted in NED01 and NED02. 
Top panel: the EFOSC2 direct B-filter image. Is drawn a ellipse-section of the ring. 
Lower panel: direct V-filter image (see \S5 for the description of the a-e letters).}
\label{fig.h}
\end{figure}

The field of HRG 2304 in Fig.~\ref{Fig.1} shows in the center three objects aligned from the N-S direction. 
The first one is a star (GSC094100228 of 14.26 V mag), the second is HRG\,2304 (NED02) and its extended 
halo-ring-like structure, and the last one is the galaxy AM\,1646-795 (NED01). After enhancing the original image by filtering processing (Fa\'undez-Abans \& de Oliveira-Abans\cite{foa98b}), Fig.~\ref{Fig.2} shows a zoom of the
field {around HRG\,2304 in the V band} to look for structures in the bulge of HRG\,2304. A projected star on 
HRG\,2304 is clearly separated from the galaxy bulge on the NW, and in the South of the C galaxy, a tinny blob is enhanced.
Bad column pixels were enhanced by the procedure. Table~\ref{table1} lists 
the stars very near HRG\,2304 and two prominent field galaxies.
In Fig.~\ref{fig.h} some structures are highlighted in NED01 and NED02; see \S5 for the description and discussion of both figures.


\begin{figure}[ht]
\begin{center}$
\begin{array}{cc}
\resizebox{\hsize}{!}{\includegraphics[]{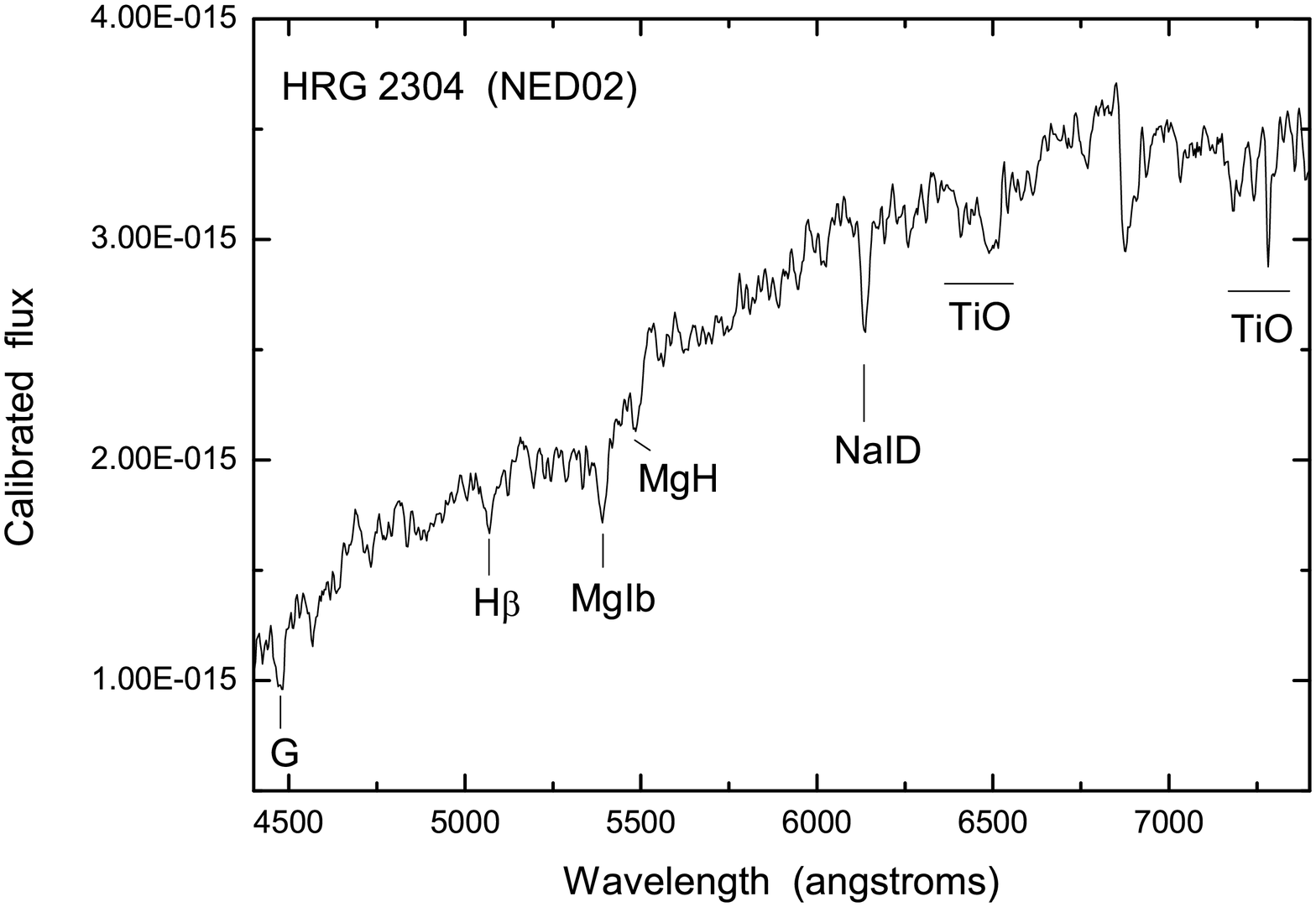}}\\
\resizebox{\hsize}{!}{\includegraphics[angle=0]{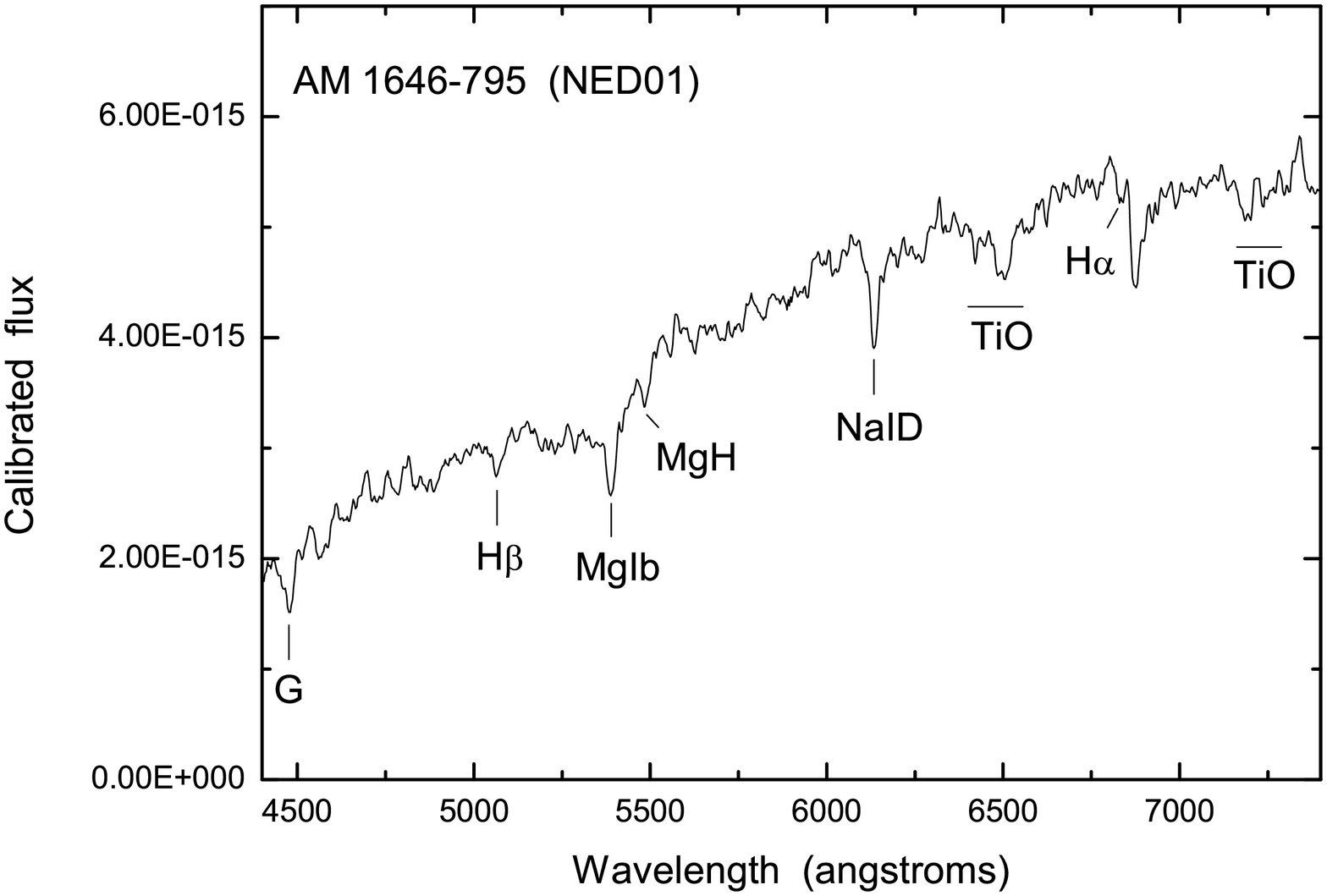}}
\end{array}$
\end{center}
\caption{Spectra of both galaxies. The vertical axis is the calibrated flux in units of 
erg  sec$^{-1}$cm$^{2}$ \AA$^{-1}$. First panel: the nuclear features of HRG\,2304. Second panel: 
same as the first panel,  but for AM\,1646-795(NED01). Overplotted in both panels is the identification of the 
main absorption features: G band $\lambda$4305\AA, H$\beta$, MgIb $\lambda$5174\AA, MgH $\lambda$5269\AA, 
NaID $\lambda$5892\AA, H$\alpha$\,and the TiO band in $\lambda$\,$\lambda$6250,7060.}
\label{fig.3}
\end{figure}


\begin{figure}[ht]
\begin{center}
$\begin{array}{cc}
\resizebox{\hsize}{!}{\includegraphics[]{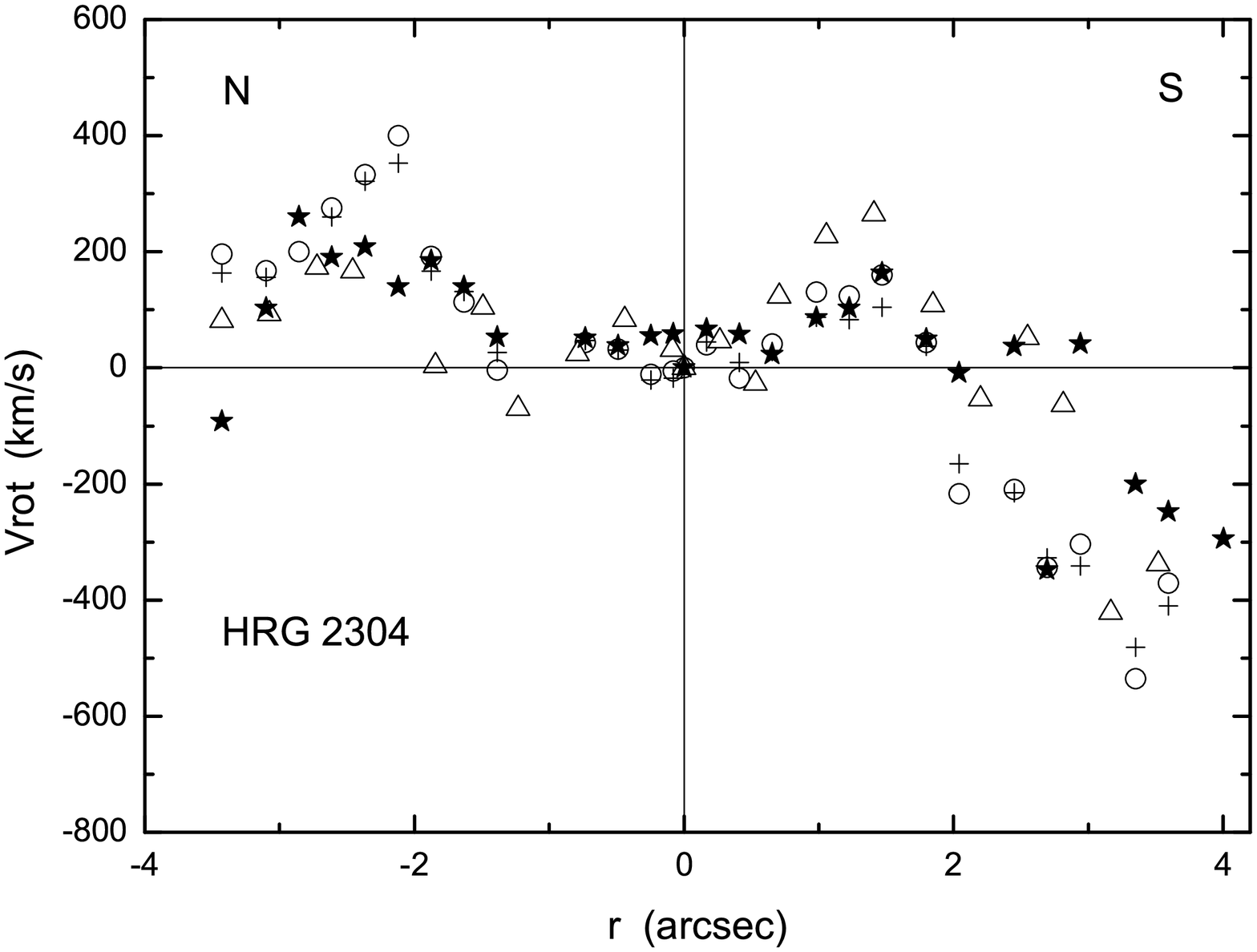}}\\
\resizebox{\hsize}{!}{\includegraphics[angle=0]{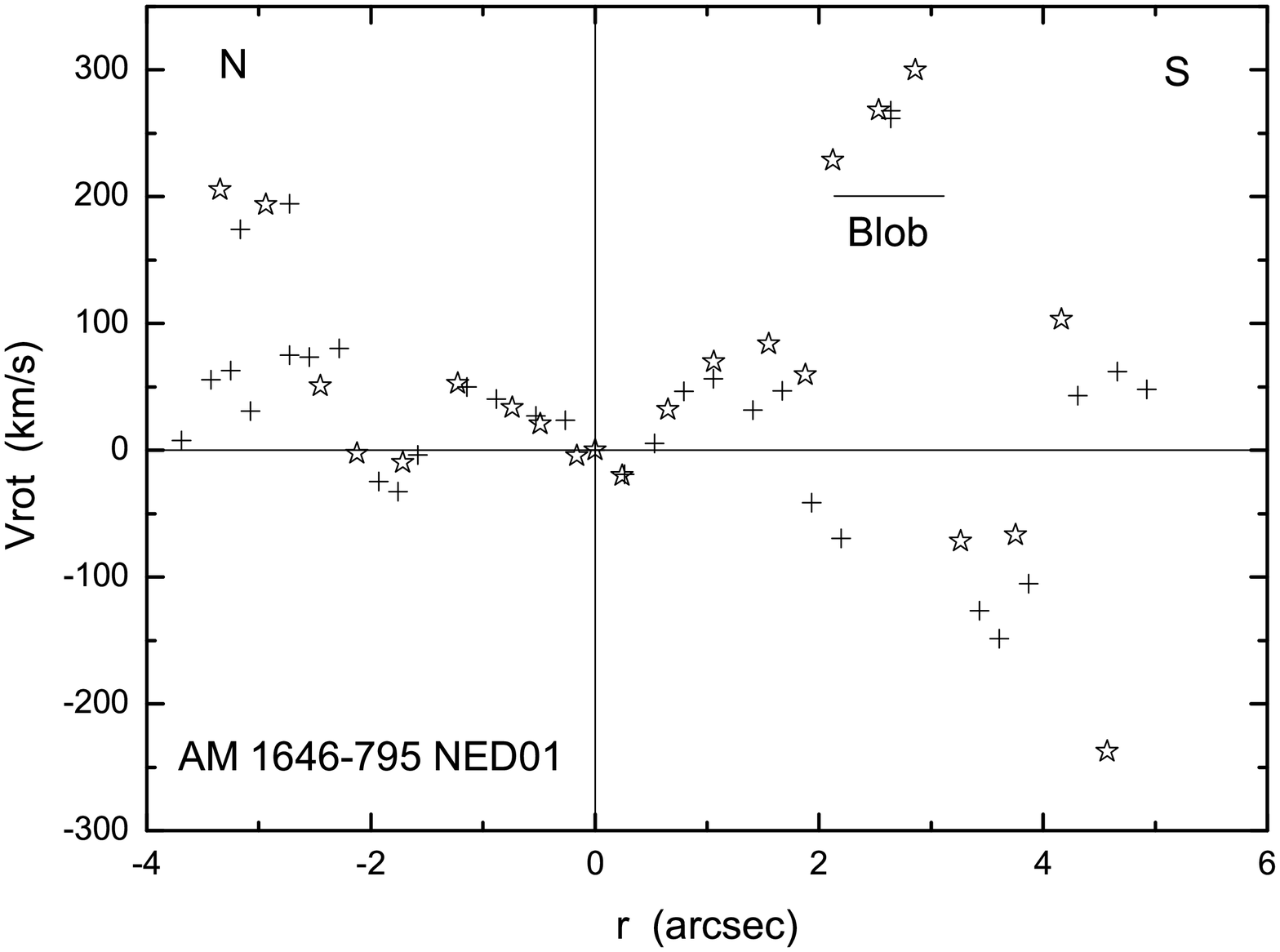}}
\end{array}$
\end{center}
\caption{The observed radial-velocity distribution for both galaxies. 
First panel is the distribution for HRG\,2304: open triangles and open circles are data from MgIb lines 
(from the two spectra, respectively); crosses are from  the NaID lines and filled stars stand for the correlated data from G, 
H$\beta$, MgIb, MgH and NaID lines. Second panel  is the same as the first one, but  for AM\,1646-795(NED01):
crosses are data from the NaID lines and open stars stand for the correlated data from G, H$\beta$, MgIb, MgH, and NaID lines.}
\label{fig.4}
\end{figure}


\begin{figure}[ht]
\begin{center}$
\begin{array}{cc}
\resizebox{\hsize}{!}{\includegraphics[]{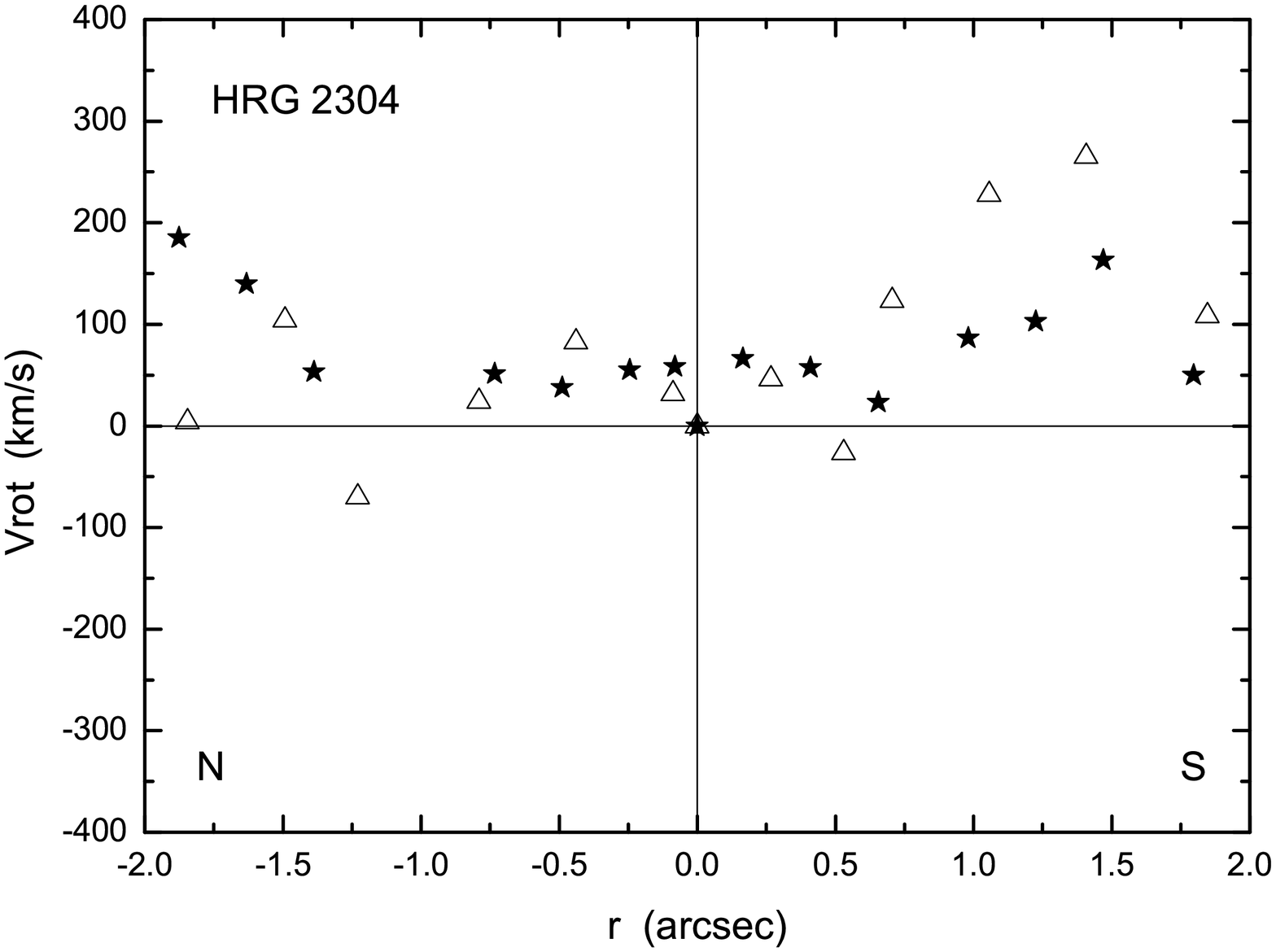}}\\
\resizebox{\hsize}{!}{\includegraphics[angle=0]{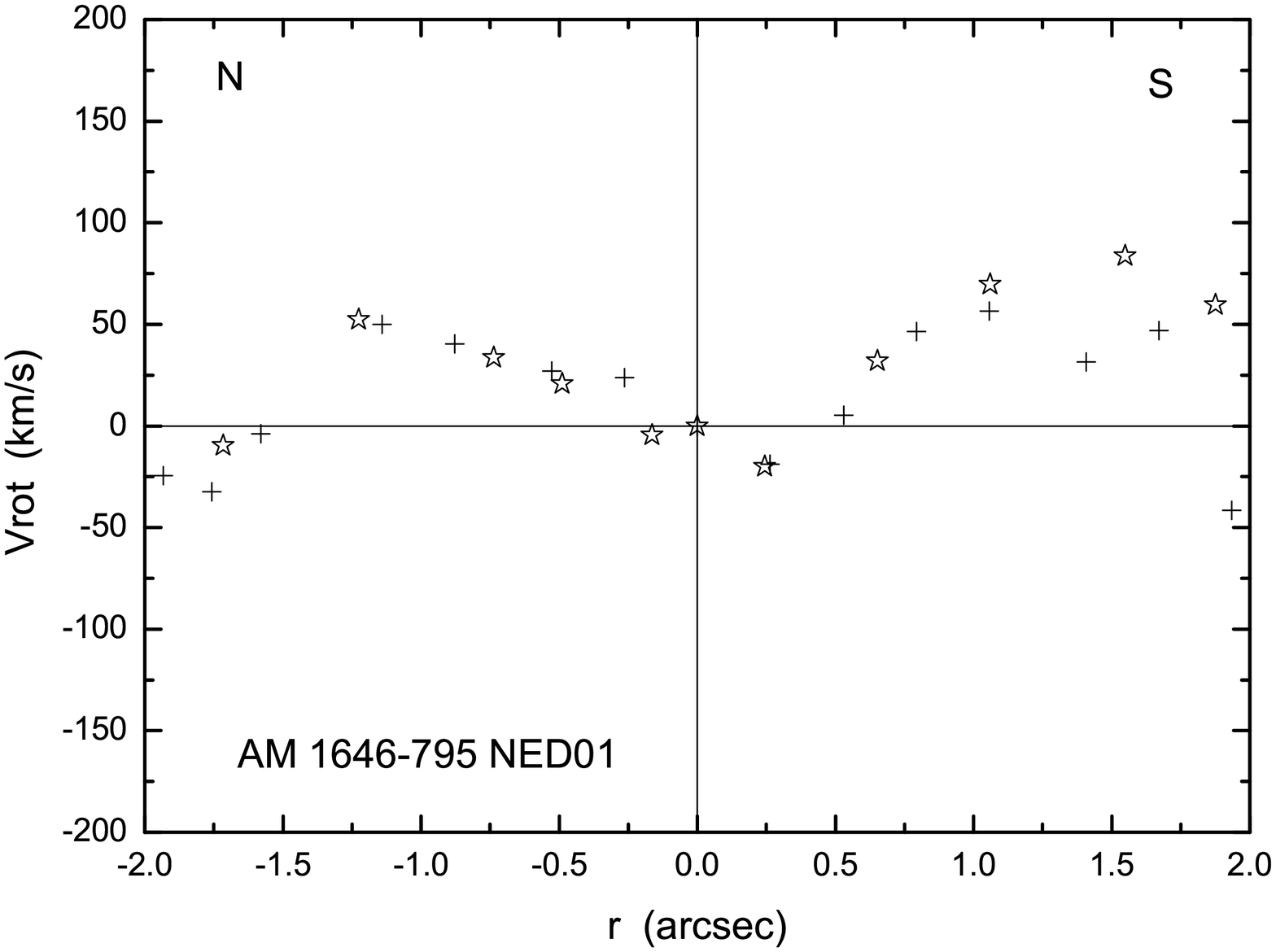}}
\end{array}$
\end{center}
\caption{The observed radial-velocity distribution for both galaxies from -2\farcs0 to 2\farcs0 
around the kinematic center. First panel is the distribution for HRG\,2304: open triangles are data from MgIb lines 
(from the two spectra, respectively); and filled stars stand for the correlated data from G, H$\beta$, MgIb, MgH 
and NaID lines. Second panel is the same as the first, but for AM\,1646-795(NED01): crosses are data 
from the NaID lines and open stars 
stand for the correlated data from G, H$\beta$, MgIb, MgH, and NaID lines.}
\label{fig.5}
\end{figure}


\begin{table*}
\caption{Stars very near to HRG\,2304 and two field galaxies.}
\label{table1}
\centering
\begin{tabular}{llll}
\hline \hline
Object & R.A. (2000) & Dec. (2000) & Remarks\\
\hline
Star & 16$^{\rm{h}}$ 54$^{\rm{m}}$ $40\fs47$ & -80\degr 03\arcmin 43\farcs98 & GSC094100288 \\
Star & 16 54 39.92 & -80 03 51.77 & The projected anonymous star on HRG\,2304\\
Galaxy & 16 54 23.47 & -80 03 53.38 & 2MASX J16542367-8003537\\
Star & 16 54 37.56 & -80 04 23.99 & Anonymous star south of C (NED01)\\
Galaxy & 16 55 30.55 & -80 05 47.92 & 2MASX J16553050-8005482 (\it a)\\
\hline
\end{tabular}
\begin{tablenotes}
\item[a]{{\it a}: Galaxy with 12\,465 km $s^{-1}$, z=0.0415, see NED.}
\end{tablenotes}
\end{table*}

\section{Observations and data reduction}

The observations were made in the service mode at La Silla Observatory 
with the ESO 3.6-m telescope equipped with ESO Faint Object Spectrograph and Camera (EFOSC2).
We used a CCD \# 40 LORAL/LESSER of 2\,048 $\times$ 2\,048 pixel, thinned and 
AR coated. A binning of 2 $\times$ 2 was adopted, implying a field on the plane of the sky  
of 2\farcm68 $\times$ 2\farcm68. The grism Gr \#4 with a grating of 
360 grooves\,$mm^{-1}$ was used. The spectra obtained covered the wavelength range of 4\,085-7\,520 \AA \,
with a scale dispersion of 1.68 \AA \,$\rm pix^{-1}$ and a FWHM resolution of 14 \AA/$1^{\arcsec}$. 
The slit width was set to 2$\arcsec$ for the galaxies  and to 5$\arcsec$
for the standard stars. Two spectra of 720 and 600 seconds were obtained. The stars used 
for extinction and flux calibration are tertiary standards from Baldwin \& Stone (\cite{bs84}), as revised by 
Hamuy et al. (\cite{h92}, see also Hamuy et al. \cite{h94}). The data reduction (bias correction, flat-fielding, cosmic ray cleaning, wavelength and flux calibrations, and 1D spectra extraction)
was made using IRAF software.

\section{Analysis and results}

We report the first optical results for the galaxies HRG\,2304 (NED02) and AM\,1646-795 (NED01) 
using long-slit spectroscopy. The new results of velocity and z, as well as some early information 
on both galaxies are listed in Table~\ref{table2}. A value of $H_{\rm o}$ = 70 km s$^{-1}{\rm Mpc}^{-1}$
was adopted throughout this work.

\subsection{The spectra and the kinematics}

The spectra of both galaxies are shown in Fig.~\ref{fig.3}, in the range 
4\,085\AA ~to 7\,520 \AA. It displays the main stellar features identified 
in the nuclear section of both galaxies: the absorption G band $\lambda$4305\AA,
H$\beta$, MgIb $\lambda$5174\AA, MgH $\lambda$5269\AA, NaID $\lambda$5892\AA, 
H$\alpha$ \,and the TiO band in $\lambda$\,$\lambda$6250,7060\AA.
Both spectra show features which are characteristic of late-type stars. 
The absorption and positions of these lines were determined by fitting a Gaussian to the observed profile. 


\begin{table*}
\caption{Data on HRG\,2304 (NED02) and AM\,1646-795 (NED01) galaxies.}
\label{table2}
\centering
\begin{tabular}{llll}
\hline \hline
Parameter &  NED01 & NED02 & Ref.\\
\hline
 R.A. (2000)  &  16 54 39.20 & 16 54 40.04 & (a) \\
Dec. (2000) & \hspace{-0.1cm}-80 04 11.00  &  \hspace{-0.1cm}-80 03 47.35  & (a) \\
pRG-family  &  & Solitaire-like & this work   \\
$z_{\rm abs}$ & 0.0415  & 0.0415 & this work \\
$V_{\rm abs}$(km\,s$^{-1}$) & 12\,430 $\pm$40 & 12\,449 $\pm$40& this work \\
$V_{\rm corr}$(km\,s$^{-1}$) & 12\,412 $\pm$90 & 12\,403 $\pm$100& this work \\
\hline
$z$  & 0.042  &   & NED \\
$V$ (km\,s$^{-1}$) & 12\,597 &  & NED\\
Other designations & 2MASX J16543910-8004109 & LEDA 228646  &NED \\
                   & 6dFJ 1654393-800411 & 2MFGC 13532 & NED \\
Magnitude & 16.5b & 11.2k & NED \\
\hline
Distance & 178 Mpc & 178 Mpc & this work \\
Mass & 1.50x$10^{11}$ M$_{\sun}$ & 0.71x$10^{11}$ M$_{\sun}$ & this work \\
Nucleus major axis ({\it B}) & 2\farcs16 & 1\farcs92 &this work \\
Nucleus minor axis ({\it B}) & 2\farcs16 & 1\farcs8 &this work\\
Bulge major axis ({\it B}) & 9\farcs0 & 7\farcs2 &this work \\
Bulge minor axis ({\it B}) & 7\farcs56 & 6\farcs72 & this work \\
Ring major axis ({\it B}) & & 15\farcs36 & this work \\
Ring minor axis ({\it B}) & & 15\farcs36 &this work \\
J - H &  & 0.277 & NED \\
H - K &  & 0.052 & NED \\
J - K &  & 0.329 & NED \\
\hline
\end{tabular}
\begin{tablenotes}
\item[a]{{\it a}: NED01 data from NED released after 20 abril 2009; NED02 data from
 Fa\'{u}ndez-Abans \& de Oliveira-Abans\cite{foa98b};\\ {\it B}: Measurements in the B-filter image.}
\end{tablenotes}
\end{table*}

For the calculation of the velocity, we considered the values corresponding to the maxima of the continuum intensity as systemic velocities. Using the MgIb and NaID absorption lines, the derived heliocentric systemic velocity $V_{\rm abs}$ for  HRG\,2304 is 12\,449 $\pm$ 90 km\,s$^{-1}$ (z = 0.0415) and for AM\,1646-795 (NED01) it is 
12\,430 $\pm$ 40 km\,s$^{-1}$ (z = 0.0415). Both galaxies form a bound system with an approximate $\Delta$V = 19 km\,s$^{-1}$. Table~\ref{table2} also displays a correlated velocity ($V_{\rm corr}$) for both galaxies 
using the G, H$\beta$, MgIb, MgH and NaID lines. We have estimated the masses based on a Virial relation employing the effective radius of each galaxy. The calculated distances and  dynamical masses for NED01 and NED02 
are 177.57 Mpc and 177.84 Mpc, 1.50x$10^{11}$ M$_{\sun}$ and 0.71x$10^{11}$ M$_{\sun}$, respectively. 
The adopted separation between both galaxy-centers is 13\farcs02, and the radius of the ring-like structure of HRG\,2304 is approximately 8\farcs7 long.

Fig.~\ref{fig.4} displays the rotation curve of both galaxies. The first panel 
shows the distribution for HRG\,2304: open triangles and circles are data from MgIb lines 
(measurements from the two spectra); crosses are from the NaID lines, and filled stars stand 
for the correlated data from G, H$\beta$, MgIb, MgH and NaID lines. The second panel shows 
the distribution for AM\,1646-795 (NED01): crosses are data from the NaID lines, and open stars 
stand for the correlated data from G, H$\beta$, MgIb, MgH and NaID lines.
Fig.~\ref{fig.5} shows the rotation curve of both galaxies, 
from -2\farcs0 to 2\farcs0 around the kinematic center.

\subsection{Stellar population synthesis}

Detailed study of the star formation in tidally perturbed galaxies not only provides important 
information on the age distribution along their stellar population components, but also helps toward understanding several aspects related to the interacting process better, along with its effects on the properties 
of the individual galaxies and their evolution.

To investigate the star formation history of HRG\,2304 and AM\,1646-795 (NED01), we use the stellar population 
synthesis code {\sc STARLIGHT}  (Cid Fernandes et al. \cite{cid04,cid05};  Mateus et al. \cite{mateus06}; Asari et al. 
\cite{asari07}). This code is extensively discussed in Cid Fernandes et al. (\cite{cid04,cid05}), and 
is built upon computational techniques originally developed for empirical population synthesis with 
additional ingredients from evolutionary synthesis models. This method was also used by Krabbe et al. (\cite{krabbe2011}) and has been successful in describing the stellar population in interacting galaxies. Briefly, the code fits an observed spectrum with a combination of $N_{\star}$ single stellar populations (SSPs) from the Bruzual \& Charlot (\cite{bruzual03}) models. These models are based on a high-resolution library of observed stellar spectra, which allows for detailed spectral evolution of the SSPs across the wavelength range of 3\,200-9\,500 \AA  ~with a wide range of metallicities. We used the Padova's 1994 tracks, as recommended by Bruzual \& Charlot (\cite{bruzual03}), with the initial mass function of Chabrier (\cite{chabrier03}) between 0.1 and 100 $M_{\sun}$. Extinction is modeled by {\sc STARLIGHT} as due to foreground dust, using the Large Magellanic Cloud average reddening law of Gordon et al. (\cite{gordon03}) with  R$_V$= 3.1, and parametrized by the V-band extinction  A$_V$. The SSPs used in this work cover 15 ages, t = 0.001\,, 0.003\,, 0.005\,, 0.01\,, 0.025\,, 0.04\,, 0.1\,,  0.3\,, 0.6\,, 0.9\,, 1.4\,, 2.5\,, 5\,, 11\,, and 13 Gyr, and three metallicities, Z = 0.2 Z$_{\sun}$, 1 Z$_{\sun}$, and 2.5 Z$_{\sun}$, adding to  45 SSP components. The fitting is carried out using  a simulated annealing plus Metropolis scheme, with regions around emission lines and bad pixels excluded from the analysis.
 
\begin{figure}
\centering
\includegraphics*[width=\columnwidth]{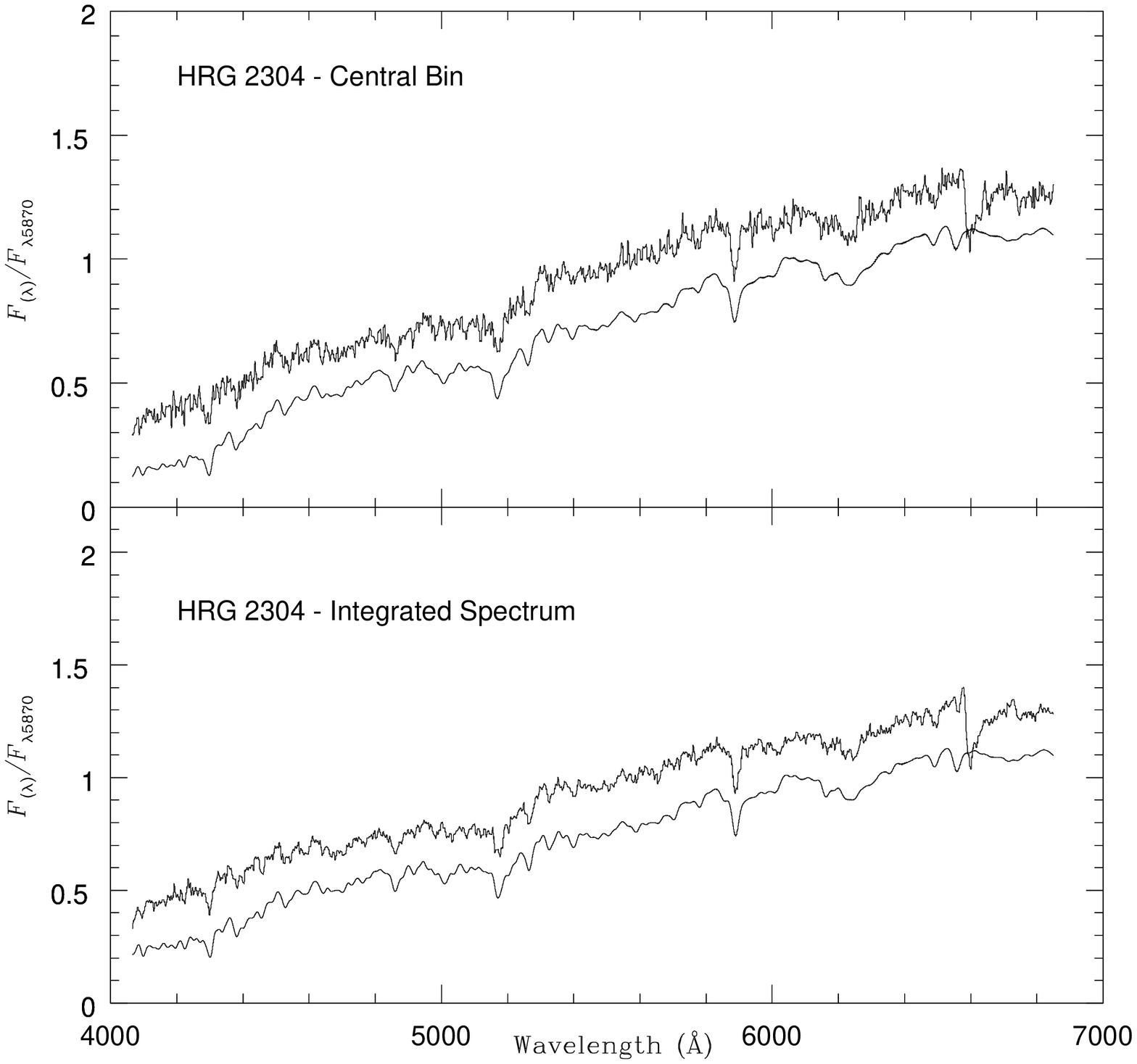}						     
\caption{Stellar population synthesis for HRG\,2304.  
Top panel: central bin spectrum corrected for reddening  and the synthesized
spectrum (shifted down by a constant). Bottom panel: same as for the top panel but for the 
integrated spectrum of the galaxy.}
\label{sintese1}
\end{figure}

\begin{figure}
\centering
\includegraphics*[width=\columnwidth]{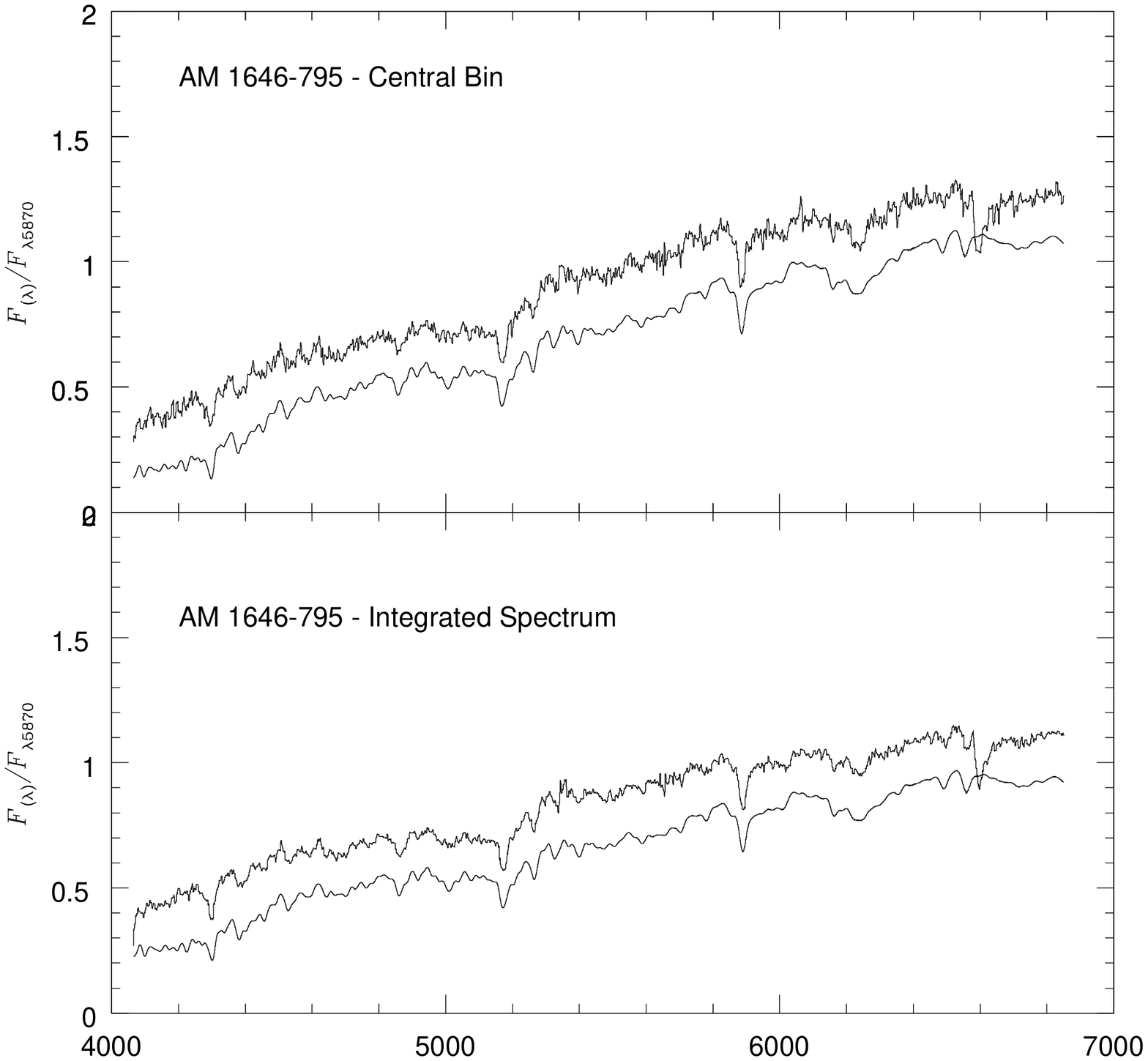}						     
\caption{Stellar population synthesis for AM\,1646-795(NED01).  
Top panel: central bin spectrum corrected for reddening  and the synthesized
spectrum (shifted down by a constant). Bottom panel: same as for the top panel but for the 
integrated spectrum of the galaxy.}
\label{sintese2}
\end{figure}

Figures \ref{sintese1} and \ref{sintese2}  show an example of the observed spectrum corrected by reddening and the
 model stellar population spectrum for HRG\,2304 and AM\,1646-795 (NED01), respectively. The results of the synthesis are summarized in Table \ref{synt_table} for the individual spatial bins in each galaxy, stated as the perceptual contribution of each base element to the flux  at $\lambda\, 5\,870$\, \AA. ~Following the prescription of Cid Fernandes et al. (\cite{cid05}), we defined a condensed population vector, 
by binning the stellar populations according to the flux contributions 
into  young, $x_{\rm Y}$ ($\rm t \leq 5\times10^{7}$ yr);
intermediate-age,  $x_{\rm I}$ ($ 5\times10^{7} <\rm t \leq 2\times10^{9}$ yr); and 
old, $x_{\rm O}$ ( $2\times10^{9} <\rm t \leq 13\times10^{9}$ yr) components. The same bins were used to
represent the mass components of the population vector $m_{\rm Y}$, $m_{\rm I}$, and
$m_{\rm O}$). The metallicity (Z), one important parameter to characterize the stellar population content,
is weighted by light fraction. The quality of the fitting result is measured by the parameters 
$\chi^{2}$ and  $adev$. The latter gives the perceptual mean deviation $|O_{\lambda} - M_{\lambda}|/O_{\lambda}$ over 
all fitted pixels, where $O_{\lambda}$ and $M_{\lambda}$ are the observed and model spectra, respectively.

\begin{figure}
\centering
\includegraphics*[angle=-90,width=\columnwidth ]{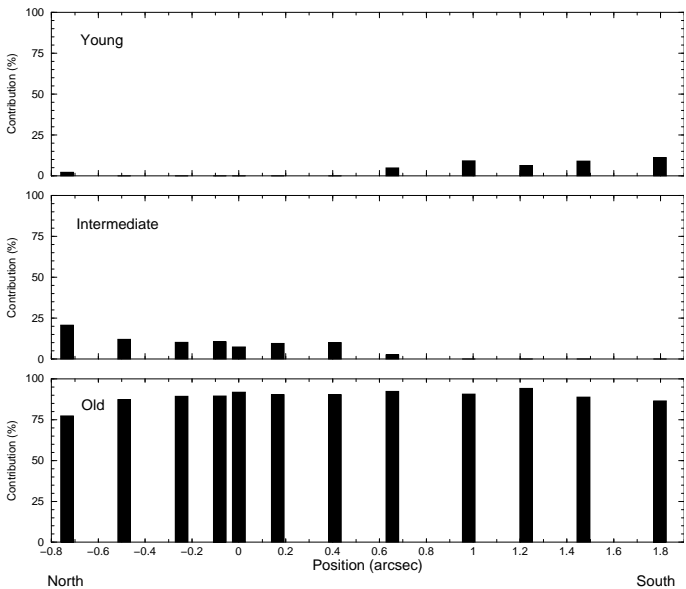}						     
\caption{Synthesis results in flux fractions as a function of the distance to the center 
of HRG\,2304.}
\label{prof_hrg2304}
\end{figure}

\begin{figure}
\centering
\includegraphics*[angle=-90,width=\columnwidth ]{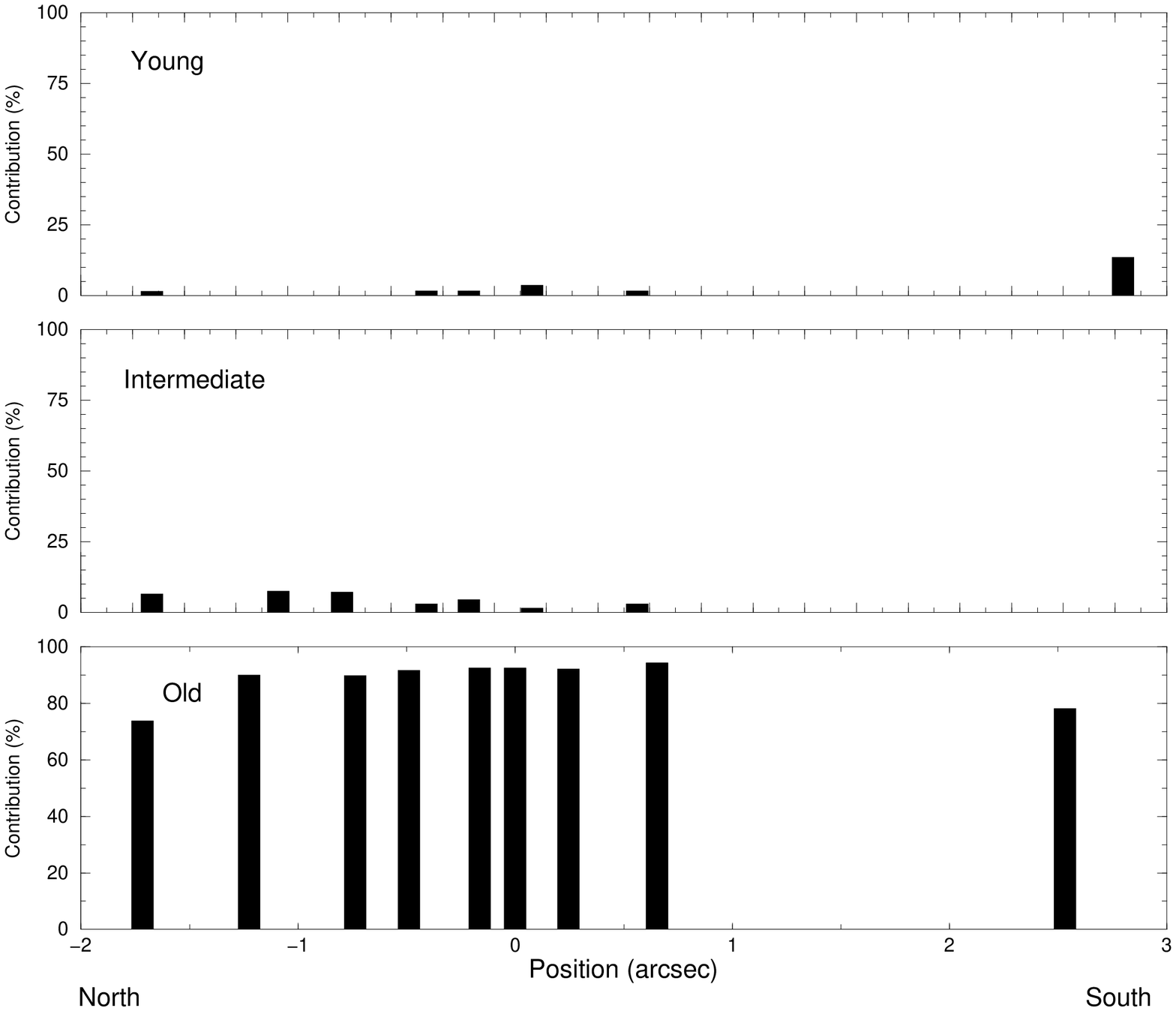}						
\caption{Synthesis results in flux fractions as a function of the distance to the center 
of AM\,1646-795(NED01)}
\label{prof_am1646}
\end{figure}

The spatial variation in the contribution of the stellar population components are 
shown in Figs.~\ref{prof_hrg2304} and \ref{prof_am1646} for the galaxies HRG\,2304 and AM\,1646-795 (NED01), respectively.
As can be seen in Figs.~\ref{prof_hrg2304} and \ref{prof_am1646}, HRG\,2304 and AM\,1646-795 (NED01)
are dominated by an old stellar population. A mean value of Z=0.028  was found for both galaxies.


\begin{table*}
\caption{Stellar-population synthesis results}
\label{synt_table}
\begin{tabular}{lrrrrrrrrrr}
 \\
\noalign{\smallskip}
\hline
\hline
\noalign{\smallskip}
Pos. &  \multicolumn{1}{c}{$ x_{\rm Y}$} &  \multicolumn{1}{c}{$x_{\rm I}$} &  \multicolumn{1}{c}{$x_{\rm O}$}
 & \multicolumn{1}{c}{$ m_{\rm Y}$}& \multicolumn{1}{c}{$ m_{\rm I}$} &
 \multicolumn{1}{c}{$m_{\rm O}$}& 
  \multicolumn{1}{c}{$Z_{\star}$[1]} &
 \multicolumn{1}{c}{$ \chi^{2}$} & 
 \multicolumn{1}{c}{$\rm adev$} & \multicolumn{1}{c}{$\rm A_{v}$}
 \\
 
\multicolumn{1}{c}{(arcsec)}& \multicolumn{1}{c}{(per cent)}  &
 \multicolumn{1}{c}{(per cent)} &  \multicolumn{1}{c}{(per cent)}
 & \multicolumn{1}{c}{(per cent)} & 
 \multicolumn{1}{c}{(per cent)}
   &
 \multicolumn{1}{c}{(per cent)}& 
  \multicolumn{1}{c}{} &
 \multicolumn{1}{c}{} & 
 \multicolumn{1}{c}{(mag)}
 \\
\hline
\noalign{\smallskip}
\noalign{\smallskip}
\multicolumn{10}{c}{HRG\,2304(NED02)}\\
\noalign{\smallskip}
\hline
\noalign{\smallskip}
-0.73 & 2.2 &  20.8  & 77.3  &  0.0& 2.9 & 97.1 &  0.036  &   1.0  &  3.25  &  0.21 \\  
-0.49 & 0.0 &  12.1  & 87.3  &  0.0& 1.0 & 99.0 &  0.026  &   1.0  &  3.13  &  0.42 \\  
-0.24 &	0.0 &  10.2  & 89.2  &  0.0& 0.8 & 99.2 &  0.027  &   1.1  &  3.36  &  0.69 \\  
-0.08 & 0.0 &  10.6  & 89.4  &  0.0& 0.8 & 99.2 &  0.027  &   1.1  &  3.24  &  0.93 \\  
 0.00 & 0.0 &  7.4   & 91.8  &  0.0& 0.6 & 99.4 &  0.029  &   1.2  &  3.78  &  1.12 \\  
 0.16 &	0.0 &  9.5   & 90.3  &  0.0& 0.7 & 99.3 &  0.027  &   1.3  &  3.43  &  1.22 \\  
 0.41 &	0.0 &  10.1  & 90.3  &  0.0& 0.8 & 99.2 &  0.028  &   1.0  &  3.49  &  1.46 \\  
 0.65 & 4.9 &  2.5   & 92.2  &  0.0& 0.2 & 99.8 &  0.028  &   1.0  &  3.68  &  1.69 \\  
 0.98 & 9.3 &  0.0   & 90.6  &  0.0& 0.0 & 100.0&  0.027  &   0.8  &  3.77  &  1.94 \\  
 1.22 &	6.4 &  0.0   & 94.1  &  0.1& 0.0 & 99.9 &  0.025  &   0.9  &  3.86  &  1.95 \\  
 1.47 &	9.1 &  0.0   & 88.7  &  0.1& 0.0 & 99.9 &  0.026  &   1.1  &  4.50  &  1.97 \\
 1.80 & 11.2&  0.0   & 86.5  &  0.1& 0.0 & 99.9 &  0.028  &   0.8  &  5.09  &  1.88 \\ 
								              
\noalign{\smallskip}
\hline
\noalign{\smallskip}
\multicolumn{10}{c}{AM\,1646-795(NED01)}\\
\noalign{\smallskip}
\hline
\noalign{\smallskip}
 -2.13 & 7.4  & 13.5 & 80.6  &  0.1&   1.9 & 98.0 &  0.031  &   0.7  &  3.76  &  0.00\\
 -1.72 & 4.9  & 21.4 & 73.8  &  0.0&   3.2 & 96.7 &  0.032  &   0.8  &  3.57  &  0.00\\
 -1.22 & 1.6  & 6.6  & 89.9  &  0.0&   0.5 & 99.5 &  0.025  &   1.1  &  3.47  &  0.00\\
 -0.74 & 0.0  & 7.6  & 89.8  &  0.0&   0.6 & 99.4 &  0.028  &   1.6  &  3.31  &  0.32\\
 -0.49 & 0.0  & 7.3  & 91.6  &  0.0&   0.6 & 99.4 &  0.029  &   1.3  &  3.35  &  0.55\\
 -0.16 & 1.7  & 3.1  & 92.4  &  0.0&   0.2 & 99.8 &  0.030  &   1.5  &  3.31  &  0.90\\
  0.00 & 1.8  & 4.5  & 92.4  &  0.0&   0.3 & 99.7 &  0.032  &   1.4  &  3.42  &  1.06\\
  0.24 & 3.8  & 1.6  & 92.1  &  0.0&   0.1 & 99.9 &  0.032  &   1.5  &  3.30  &  1.30\\
  0.65 & 1.7  & 3.0  & 94.3  &  0.0&   0.4 & 99.6 &  0.019  &   1.5  &  4.45  &  1.84\\
  2.53 & 13.6 & 0.0  & 78.1  &  0.2&   0.0 & 99.8 &  0.020  &   0.6  &  6.51  &  2.20\\
\noalign{\smallskip}
\hline
\noalign{\smallskip}
\noalign{\smallskip}
\noalign{\smallskip}
\end{tabular}
\begin{minipage}[c]{18.0cm}
[1] Abundance by mass with Z$_{\sun}$=0.02 \\
\end{minipage}
\end{table*}


\section{Discussion}

Along all the slit, the spectra of both galaxies show features which are characteristic of late-type stars. 
No star-forming regions and no nuclear ionization sources were detected. The two galaxies are a tidally bound 
system with a radial velocity difference of almost 19 km\,s$^{-1}$. There are two  nearby 
galaxies in the field: (a) 2MASX J16542367-8003537 without redshift data in the literature, and 
(b) 2MASX J16553050-8005482 with z=0.0415, 12\,465 km\,s$^{-1}$. 
In this context, HRG\,2304 and AM\,1646-795 (NED01) are members of a nearby group of galaxies, and probably all the components are gravitationally bound. The HRG\,2304 and AM\,1646-795 (NED01) system is at a distance of about 178 Mpc. From the quoted mass, HRG\,2304 has approximately half of the mass of AM\,1646-795 (NED01).

The errors of individual velocity measurements do not exceed 20 km\,s$^{-1}$ in the central region of the galaxies and increase to 30-50 km\,s$^{-1}$ on its periphery. There is a significant dispersion on the radial velocity distribution of HRG\,2304 (see first panel of Fig.~\ref{fig.4}). The velocity dispersion are $\pm$40 km\,s$^{-1}$ around the position of $\pm$1\farcs0, and $\pm$90 km\,s$^{-1}$ between a radius $\pm$2\farcs0 to $\pm$4\farcs0. With respect 
to the kinematic center, the north region is receding from us, while the one in the south, 
waved into receding (between a radius of $\sim$   0\farcs1 to 2\farcs0) and approaching us
(between a radius of $\sim$ 2\farcs0 to 4\farcs0). As in the first panel of Fig.~\ref{fig.4}, the radial velocity distribution of AM\,1646-795 (NED01) varies in almost the same way as HRG\,2304. There is an object tight-south to the very central region, which was detected by the luminosity profile in the spectral extraction 
(the Blob in the second panel of Fig.~\ref{fig.4}). The latter is receding from us and its velocity 
suggests that it is a satellite bound to NED01.

Inspecting Fig.~\ref{fig.5}, and following the variation of the radial velocity distribution it may be suggested that: (a) in the first panel, the nuclear region is approaching us; and (b) in the second panel, 
the nuclear region is receding from us. In both, the galaxy components from -2\farcs0 to 2\farcs0 
around the kinematic center are receding from us.   

The stellar formation history of both galaxies were well extracted by the stellar population synthesis code STARLIGHT 
(see Figs.~\ref{sintese1} and ~\ref{sintese2}). The synthesis results in flux fraction as a function of the distance 
to the center of each galaxy show that the spatial variation in the contribution of the stellar-population components 
in both objects are dominated by an old stellar population with ages between $2\times10^{9} <\rm t \leq 13\times10^{9}$ yrs. 
The latter confirms that both galaxies are early-type objects (elliptical-like galaxies and/or poor-gas S0 galaxies with an 
elliptical companion). The stellar population synthesis indicated that there is a small, but non-negligible, fraction of 
young stars (see first panel of Figs.~\ref{prof_hrg2304} and ~\ref{prof_am1646}). Those young stars are principally 
in HRG\,2304 in the direction of the ``contact" region between both galaxies, the S-region in HRG\,2304 and 
the N-region in AM 1646-795 (NED01). Studies based on absoption line spectroscopy and studies of the fundamental plane 
favor a ``frosting" model in which early-type galaxies consist of an old base population with a small amount of younger 
stars (Trager et al. \cite{trager2000}; Gebhart et al. \cite{gebhardt2003}; and Schiavon \cite{schiavon2007}). Besides, 
data on early-type galaxies which exhibit strong ultraviolet excess, polycyclic aromatic hydrocarbon emission and infrared 
excess, are interpreted as a possible result of recent low-level star formation (Yi et al. \cite{yi2005}; Rich 
et al. \cite{rich2005}; Schawinski et al. \cite{scha2007}; Kaviraj et al. \cite{kaviraj2007}; Temi et al. \cite{temi2009}; 
Young et al. \cite{young2009} and Salim \& Rich \cite{sr2010}). Studies based in spatially resolved spectroscopy by 
Shapiro et al. (\cite{sha2010}) and Kuntschner et al. (\cite{kun2010}) find that star formation in early-type galaxies 
happens exclusively in fast-rotating objects and occurs in two different contexts: (a) objects with widespread young
 stellar population associated with a high molecular gas content; (b) objects with disk and/or ring morphology. The 
latter could be explained by rejuvenation in previously quiescent stellar systems (Shapiro et al. \cite{sha2010}; 
Kuntschner et al. \cite{kun2010}), and seems to be suitable to explain HRG\,2304. 

In spite of the lack of results between 1-2 arcsec in the lower panel of Fig.~\ref{prof_am1646}, the old stellar population distribution appears homogeneous for both galaxies (see also Figs.~\ref{prof_hrg2304}). The intermediate population appears to be asymmetrically distributed in both galaxies (see the middle panel of Figs.~\ref{prof_hrg2304} and ~\ref{prof_am1646}). HRG\,2304 seems to be different from other morphologically-similar objects which show nuclear activity and recent burst of star formation in the ring. As an example, the apparently morphologically-related ring galaxy NGC 985 shows Seyfert 1 activity, and the observations support the hypothesis that it contains two closely spaced nuclei as a result of the collision of two galaxies: a disk galaxy with an elliptical/spheroidal one, both or one of them with significant molecular and dust content. Different from HRG\,2304, the ring galaxy NGC 985 is an object which has been extensively studied because of its peculiar characteristics (see NED; Appleton \& Struck-Marcell \cite{app96}; Arribas et al. \cite{arribas1999}; Appleton et al. \cite{app2002}; Krongold et al. \cite{kron2005}, among others. On the other hand, HRG\,2304 seems to be more closely related to the ring galaxy FM 47-02, an early-type system which seems to originate in the tidal interaction of a probable E/S0 galaxy with an elliptical E4 companion (Fa\'{u}ndez-Abans et al. \cite{fa_oa2010}; \cite{fa_k2011}).

The ring-like structure of HRG\,2304 was probably triggered by a wave of enhanced density moving outwards, resulting only in the redistribution of the old stellar population of this galaxy. This structure resembles the asymmetric distribution of the galaxy contents seen in the simulations of slightly off-center collisions of Appleton \& Struck-Marcell (\cite{app_st1987}).

The roughly-derived ellipticity of the ring-like structure of HRG\,2304 is 0.16 with an adopted radius of 7.5 kpc.
There are some structures inside and around the ring, with a few of them marked by ``a" in the lower panel of Fig.~\ref{fig.h} (those are smooth clumps). We have also highlighted: the bright arc section of the ring ``b"; the distorted bulge region of AM\,1646-795 (NED01) ``c", in the direction of HRG\,2304; the fanlight structure ``e", 
associated with AM\,1646-795 (NED01); and a bright satellite candidate ``d" (a more compact structure). 
In the first panel of Fig.~\ref{fig.h}, a section of the ring-like structure encompassing a
few of its principal components is drawn. The related structures together with the ring around HRG\,2304 are 
evidence of tidal interaction of both objects. The morphological and late-type star spectral characteristics, 
dominated by old stellar population in both objects and the off-center nucleus of HRG\,2304 in the 
direction of NED01 (off-sets: 1\farcs92 N$\rightarrow$S and 0\farcs48 W$\rightarrow$E direction), 
suggest that HRG\,2304 can be proposed as candidate of Solitaire ring galaxy in its early stages 
of interaction with the companion.

\section{Conclusions}

In this work, we report observations of the peculiar Ring Galaxy HRG\,2304, 
which was previously classified as an elliptical-type ring galaxy by FAOA. 
Our work is based on low-resolution spectroscopic observations in the optical band to highlight some spectral characteristics of this object and its companion AM\,1646-795 (NED01).

In summary, the results of our work on HRG\,2304 are:

\begin{itemize}

\item The spectra of both galaxies show features that are characteristic of late-type stars. No star-forming regions and no nuclear ionization sources were detected.

\item The calculated heliocentric systemic velocity for HRG\,2304 is 12\,449 $\pm$ 40 km$s^{-1}$ ({\it z} = 0.0415), and for AM\,1646-795 (NED01) it is 12\,430 $\pm$ 40 km$s^{-1}$ ({\it z} = 0.0415); the latter is in agreement with early values in the literature. 

\item The calculated masses are 0.71x$10^{11}$ M$_{\sun}$ for HRG\,2304 and 1.50x$10^{11}$ M$_{\sun}$ for AM\,1646-795 (NED01), meaning that HRG\,2304 has approximately half the mass of AM\,1646-795 (NED01). Both objects are in a close encounter, and the separation between both galaxy centers is 13\farcs02. The tidal interaction triggered the formation of a ring-like structure in HRG\,2304 with a radius of approximately 8\farcs7 long.

\item Both galaxies are at the same distance on the sky plane, and HRG\,2304 seems to be face-on.

\item In both galaxies, the spatial variation in the contribution of the stellar-population components are dominated by an old stellar population of $2\times10^{9} <\rm t \leq 13\times10^{9}$ yr.

\item There is a little, but non-negligible, fraction of young stars which is evident after the extraction by the stellar population synthesis code STARLIGHT. This suggests a probable rejuvenation in previously quiescent stellar systems remaining in both galaxies.

\item The ring-like structure of HRG\,2304 was probably triggered by the tidal wave moving outwards, resulting only in the redistribution of the old stellar population of this galaxy. The latter suggests that HRG\,2304 did not have enough gas and dust to trigger star formation in regions compressed by the density wave.

\end{itemize}

Finally, the goal of this paper is to bring the first input in the study of the Solitaire-class of Ring Galaxies. The spectroscopic results and the morphological peculiarities of HRG\,2304 and its companion AM\,1646-795 (NED01), can be adequately interpreted as an ongoing stage of interaction. Both galaxies are early-type, the companion is elliptical, and the smooth distribution of the material around HRG\,2304 and its off-center nucleus in the direction of AM1646-795 (NED01) characterize HRG\,2304 as a candidate to Solitaire-type Ring Galaxy in its early-stage of ring formation. Despite the low resolution of the direct images and the uncertainties related to the possible group companionship and dynamics, we intend to conduct an exploratory study by employing numerical N-body/hydrodynamical simulation to reconstruct this history and to predict the evolution of the tidal interaction. This exploration of collision parameters will certainly contribute to a better understanding of the Solitaire forming process, probably in the sense of constraining the group of physical parameters and boundary conditions.

\begin{acknowledgements}
      
This work was partially supported by the Brazilian Minist\'{e}rio da Ci\^{e}ncia e Tecnologia (MCT), 
Laborat\'{o}rio Nacional de Astrof\'{i}sica (MCT/LNA), and Universidade do Vale do Para\'{i}ba - UNIVAP. A. C. Krabbe  thanks the support of FAPESP, process 2010/01490-3. We also thank Ms. Alene Alder-Rangel for editing the English in this manuscript. This research made use of the NASA/ IPAC Infrared Science Archive, which is operated by the Jet Propulsion Laboratory, California Institute of Technology, under contract with the National Aeronautics and Space Administration. 

\end{acknowledgements}

\end{document}